\documentstyle[pre,aps,multicol,epsf]{revtex}
\begin{document}
\draft
\title{Anomalous Dynamics of Translocation}
\author{Jeffrey Chuang$^1$, Yacov Kantor$^2$, and Mehran 
Kardar$^{1,3}$}
\address{
$^1$Department of Physics, Massachusetts Institute of Technology,
Cambridge, Massachusetts 02139 \\
$^2$School of Physics and Astronomy, Tel Aviv University, Tel Aviv 69 
978, 
Israel\\
$^3$Institute for Theoretical Physics, University of California,
Santa Barbara, California 93106\\
}
\maketitle
\begin{abstract}
We study the dynamics of the passage of a polymer through a membrane 
pore
(translocation), focusing on the scaling properties
with the number of monomers $N$. 
The natural coordinate for  translocation is the 
number of monomers on one side of the hole at a given time.
Commonly used models which assume Brownian dynamics for this variable 
predict a mean (unforced) passage time $\tau$ that scales as $N^2$,
even in the presence of an entropic barrier.
However, the time it takes for a free polymer to diffuse a distance of 
the order of its radius by Rouse dynamics scales with an exponent 
larger 
than 2, and this should provide a lower bound to the translocation 
time. 
To resolve this discrepancy, we perform numerical simulations with 
Rouse dynamics for both phantom (in space dimensions $d=1$ and 2), 
and self-avoiding (in $d=2$) chains. 
The results indicate that for large $N$, translocation times 
scale in the same manner as diffusion times, but with a larger 
prefactor 
that depends on the size of the hole. 
Such scaling implies anomalous dynamics for the translocation process.
In particular, the fluctuations in the monomer number at the hole are
predicted to be non-diffusive at short times, while the average pulling
velocity of the polymer in the presence of a chemical potential 
difference
is predicted to depend on $N$.
 
\end{abstract}

\begin{multicols}{2}
\narrowtext

\section{Introduction}
The process of translocation, in which a polymer worms its way
through a narrow pore, is an event important to many biological
systems. Examples include the viral injection of DNA into a host, 
DNA packing into a shell during
viral replication, gene swapping through bacterial pili, and
the genetic technique of cell transformation
by DNA electroporation\cite{MolBioCell}.
There are also a number of recent {\em in vitro} experiments on 
translocation, e.g. the electric field-induced migration  of DNA through
microfabricated channels\cite{han} or through an $\alpha$-hemolysin 
protein channel in a 
membrane\cite{Kasianowicz,meller}.
The driving force is an essential ingredient in the above process,  
as are the entropic and cooperative factors that 
arise from the connectivity of the polymer.
An interesting statistical consequence of the latter is that the 
polymer faces an entropic barrier, since the number of available 
configurations
is least when the chain is halfway through the hole.
In this regard it shares similarities with other entropically 
controlled
polymer systems, e.g. polymer trapping 
in random environments \cite{entrap1,entrap2,entrap3},  DNA gel 
electrophoresis\cite{electrophor} or reptation \cite{deGennes_book}. 
In these cases, the geometry of the obstacles around which the polymer 
must diffuse constrains the kinetics of the process.

A number of recent theoretical works have shed light on the 
translocation
process\cite{Kasianowicz,muthu,muthuapril2001,deGennesDNA,Lubensky,bs},
mostly in the presence of a driving force.
A common approach is to focus on the dynamics of a single
variable representing the monomer number at the 
pore\cite{muthu,Lubensky,park}.
Due to its resemblance to the `reaction coordinate' for chemical 
processes,
we shall refer to this parameter as the translocation coordinate.
Assuming that the segments on the two sides of the hole are in 
equilibrium 
leads to a force acting on the trapped monomer which can be derived 
from the 
entropic barrier mentioned before, as well as any chemical potential
differences that may provide a driving force. 
The translocation problem is thereby reduced to the escape of a 
`particle'
(the translocation coordinate) over a potential barrier.

\begin{figure}[htb]
\epsfysize=15\baselineskip
\centerline{\hbox{ \epsffile{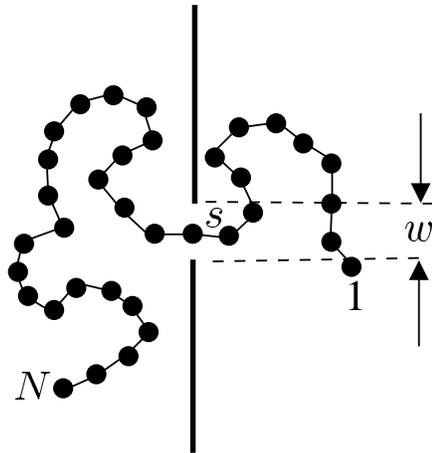}}}
\caption {\protect Schematic representation of a $N$--monomer polymer
in the process of translocation through a hole of size $w$.
The `translocation coordinate' $s$ is the number of monomers on one 
side.}
\label{transloc}
\end{figure}

Assuming {\em Brownian dynamics} for the translocation coordinate, 
and in the absence of a driving force, 
the characteristic first passage time scales as $N^2$, 
where $N$ is the number of monomers.
This result cannot be reconciled with the equilibration time of a 
polymer
which  scales as $N^{\alpha}$, with a dynamic scaling exponent $\alpha$
that is larger than 2 for Rouse dynamics of self-avoiding chains. 
Clearly, we expect the constraint of passage through a hole 
to slow down rather than speed up the dynamics of the polymer,
conceivably leading to an exponent even larger than $\alpha$.
Previous work recognizes this difficulty; for example, 
Ref.\cite{Lubensky}
while using the Brownian particle analogy, clearly explains why it 
is applicable to the pre-asymptotic region of interest in that paper.
In this work we consider the true asymptotic scaling of the 
translocation 
time for large $N$, and find that its scaling is similar to the 
corresponding
equilibration time, albeit with a larger prefactor.
Since we cannot rely on the Brownian particle picture in this regime, 
we reach
this conclusion by numerical simulations.

The problem is to calculate the time required for a polymer to move 
from one side of a rigid wall to the other through a narrow hole. 
This is schematically depicted in Fig.~\ref{transloc}, with the space 
on both sides of the wall being infinite. 
Although frequently a driving force, such as an external field or
chemical potential difference is present in the problem, we shall 
restrict ourselves to a model without external forces. 
For translocation to occur, there must be two events. The first is the 
collective diffusion of the polymer to the vicinity of the pore; 
the second is its threading through the pore. 
For a finite system, or in the presence of a finite concentration,
the first event takes a time determined by the concentration of 
polymers in solution, their diffusion constant, and the
effective cross-section of the hole.
This time is decoupled from the time for the second event, which is
constrained by the passage of all monomers through the hole.
Since we are interested only in the latter event, we shall  
assume that in the initial state the first monomer of the polymer chain 
is already threaded through the hole. 
To avoid the situation in which the polymer withdraws from the hole
and drifts away to infinity, we  add the restriction 
that the first monomer is never allowed to cross back out of the hole. 
These constraints effectively isolate 
the translocation time from the time for the polymer to find the hole
\cite{muthuapril2001}.

Despite the conceptual simplicity of the translocation problem, 
it has been difficult to solve analytically.
Even the simplified case of a Gaussian polymer in a one--dimensional 
space 
moving past a potential barrier is non--trivial\cite{onedbar}.
Consequently, theoretical treatments resort to approximations such
as reducing the problem to Brownian dynamics of the translocation 
coordinate.
As reviewed in Sec.~\ref{analysis}, the focus of this approach is the
probability density function $p(s,t)$ that a particular monomer 
(labeled
by its sequential number $s$ along the chain) is located at the hole at 
time 
$t$. For such a constrained configuration, one can derive the entropy 
of the 
polymer if the segments on the two sides are in equilibrium.
This entropy is then assumed to generate a force acting on the monomer,
favoring its motion to one side or the other.
Naturally, stochastic forces are also present (and are in 
fact necessary to push the chain over the entropic barrier).
Assuming that the translocation coordinate obeys Brownian dynamics
leads to  a Fokker-Planck equation for the evolution of $p(s,t)$.
The standard Kramers' approach to escape over a potential barrier 
yields
a mean translocation time that scales with the number of monomers as 
$N^2$,
i.e. the entropic barrier does not modify the diffusive scaling.
We supplement this result with a numerical integration of the 
Fokker-Planck
equation that yields the complete distribution function for transit 
times.
The Rouse model for the dynamics of a {\em phantom chain\/} also 
predicts a 
time of order of $N^2$ for the equilibration of the polymer.
Since such equilibration is essential to the use of the entropy
function as the driving force, the internal consistency of the approach 
is in doubt.
This is even more so when considering self-avoiding polymers, where the
relaxation times are much larger than $N^2$.
Noting this contradiction we proceed to numerical simulations in the 
rest
of the paper.

In Sec.~\ref{sim1d} we report on simulations of a phantom chain in
one dimension. In this case the relaxation time of the chain scales as
$N^2$, and is thus not inconsistent with the predictions of the 
Brownian
particle analogy. We do indeed find that the probability density 
function
for translocation times (once appropriately scaled) is quite similar to
that obtained from solving the Fokker--Planck equation.
However, the mean translocation time, while appearing to scale as 
$N^2$,
has a much larger prefactor, which depends on the size of the hole.
In one dimension self-avoiding chains are fully stretched,
and do not provide a fair model for translocation of a coiled polymer.

Two dimensional polymers as described in Sec.~\ref{sim2d}, are ideal 
for
studying the scaling of translocation times in a more complex 
situation.
Simulations are faster and easier than in three 
dimensions,
while at the same time the effects of self-avoidance are more 
pronounced.
Our simulations for phantom chains reproduce the trends observed in
one dimension, i.e. a mean translocation time scaling as $N^2$ with a 
larger
prefactor. However, once self-avoidance is included the translocation
times increase dramatically, the mean translocation time appearing to
scale as $N^\alpha$, with $\alpha\approx 2.5$ which is the exponent for 
Rouse relaxation of self-avoiding chains in two dimensions.
We find that (for the parameters of our model)  translocation times 
are roughly  ten times longer than typical equilibration times. 
Thus translocation is indeed much slower than diffusion of the polymer,
but appears to scale with the same exponent.

Consequences of this observation are discussed in 
Sec.~\ref{discussion}.
The observed nontrivial scaling of translocation times is clear 
indication
of the failure of the Brownian picture for the dynamics of the 
translocation
coordinate.
Instead, we suggest that the anomalous dynamics of a specified monomer 
in a chain provides a better analogy.
Following scaling arguments used for the latter, we predict anomalous
behaviors for the translocation coordinate $s(t)$. 
In particular, fluctuations in $s$ are predicted to scale as $t^\zeta$,
while in the presence of a chemical potential $\Delta\mu$, the pulling
velocity is predicted to behave as $u\propto \Delta\mu N^\eta$, with
$\zeta\approx0.46$ and $\eta\approx -0.18$ for Rouse dynamics in three
dimensions.

\section{Brownian Translocation}\label{analysis}

The reduction of the translocation problem to the Brownian dynamics of 
a
single coordinate was introduced in Ref.\cite{sung}, and further 
explored in Ref.\cite{muthu}. 
Here we review the main features of this approximation
and its consequences.
Consider a polymer moving through a pore in a membrane, where the hole 
is
so narrow that only a single strand of polymer can pass through.
(Thus, the parameter describing the width of the hole in 
Fig.~\ref{transloc}
is $w=1$.)
The progress of the polymer can be tracked by following the
number $s$ of the monomer which is located in the hole at a particular 
time,
as depicted in Fig.~\ref{transloc}. 
Let us denote the probability of monomer $s$ being in the 
hole at time $t$ by $p(s,t)$.
As the monomer $s$ moves forward or backward through the hole a 
distance $a$ 
(of order of the typical separation between monomers) the relevant 
monomer 
number increases or decreases by unity.
Treating $s$ as a continuous variable, we can  write a continuity 
equation
for the probability as
\begin{equation}
{\partial p\over \partial t}+{\partial j\over\partial s}=0 ,
\label{Smol}
\end{equation}
where $j(s,t)$ is the probability current.

The central difficulty is to find an appropriate expression for $j$ 
which 
correctly reflects the correlated motion of the whole polymer.
If the progress of the polymer  is sufficiently slow for the
segments on the two sides to come to equilibrium, the monomer at the 
hole
experiences a mean force that can be obtained from
the variations of a constrained free energy $F(s)$.
How such a force can be used to deduce the dynamics of the monomer 
label 
is not clear. 
The analogy to Brownian motion suggests that the rate of change of $s$ 
is 
related to the force by a mobility $\mu$. 
Since the polymer fluctuates back and forth between the two sides, 
there 
must also be a stochastic element which can be represented by a random 
force.
If there are no correlations in this force at different times (as in 
the standard Langevin formulation for a Brownian particle),
there is a current that depends
on the local probability density as
\begin{equation}
j=-D\left( {\partial p\over\partial s}+
{p\over k_BT}{\partial F\over\partial s}\right)\ ,
\label{jdef}
\end{equation}
where the diffusion parameter $D$ is  proportional to the variance 
of the stochastic force\cite{sung-comment}.
As in standard Brownian processes, the above equation assumes that the
mobility is related to the temperature $T$ by $\mu=D/k_BT$, where $k_B$
is the Boltzmann constant.
No similar restriction is made in Ref.\cite{Lubensky}, which obtains 
the 
diffusion term from considerations of symmetry and locality.
However, as elaborated in Sec.~\ref{discussion}, the assumption of 
locality 
need not be valid in this case, since the true dynamics of $s$ must 
reflect the collective behavior of the whole polymer.

\begin{figure}[htb]
\epsfysize=15\baselineskip
\centerline{\hbox{ \epsffile{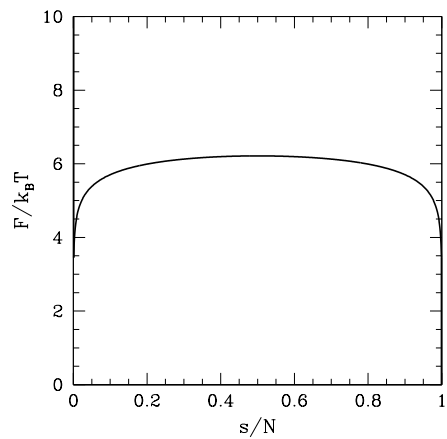}}}
\caption {\protect The entropic  potential barrier found in the single 
variable equivalent to the polymer translocation problem.}
\label{potential}
\end{figure}

Calculating the restricted free energy $F(s)$ is reasonably 
straightforward,
and equivalent to finding the number of possible configurations of a 
polymer 
attached at one point to an impenetrable barrier.
The exact solution is known for the case of one--dimensional  discrete 
random walks with fixed  step--length (see, e.g., Ref.~\cite{chandra}): 
It can be shown
that (in the large $N$ limit) the number of $N$-step walks which start
at a boundary and never return to it is $\sqrt{2/\pi N} \cdot 2^N$. 
Thus, the number of configurations with $s$ monomers on the right and 
$N-s$ monomers on the left has the $s$--dependence $A/[(N-s)s]^\gamma$, 
where 
$A$ is independent of $s$ and $\gamma={1/2}$, giving 
the $s$--dependent part of the free energy as 
\begin{equation}
F=\gamma k_BT\ln[(N-s)s] .
\label{FreeEn}
\end{equation}
To this result we add the conditions that the first monomer can
never be withdrawn from the hole, and that after the $N$th monomer
crosses the wall the polymer will no longer return to it.
Figure~\ref{potential}
depicts the resulting free energy for the case of $N=1000$. The two 
conditions
are shown by two vertical lines on the sides of the graph: 
the line on the left (infinite barrier)
signifies our assumption that the first monomer can never cross back
through the hole, while the vertical line on the right ($-\infty$)
represents the escape of a polymer which has crossed the barrier.
The numbers of configurations of polymers in higher space dimensions 
$d$, 
whether phantom or self-avoiding (SA), cannot be calculated exactly. 
However, it is known that they have
the same dependence on the polymer length\cite{expon}, with  $\gamma$ 
in Eq.~(\ref{FreeEn}) replaced by a number which 
depends on $d$ ($\gamma=d/2$ for phantom chains). Thus the $s$ 
dependent part of the free energy will have a different prefactor, 
but the logarithmic dependence will remain unchanged.

By inserting Eq.~(\ref{FreeEn}) into  Eqs.~(\ref{Smol}) and 
(\ref{jdef}), 
we reduce polymer translocation to a  standard 
single particle problem of escape over a potential barrier. It is 
interesting
to note that if we rescale the variables according to $s\rightarrow 
sN$,
$t\rightarrow tD/N^2$, all $N$-- and $D$--dependence is eliminated
from the equation, resulting in
\begin{equation}
{\partial p\over \partial t}={\partial^2 p\over\partial s^2}+
\gamma{\partial\over\partial s}
\left(  {1-2s\over (1-s)s}p\right) .
\label{stdeqn}
\end{equation}
The solution to this dimensionless equation can be converted back to 
real time 
by
multiplying the $t$ axis by $N^2/D$. Thus, under the assumptions listed
above, the escape time of a particle, and thus the translocation time
of a polymer are proportional to $N^2/D$. Note, that this conclusion is
{\em independent} of the value of the parameter $\gamma$, and
remains valid even for a SA chain in which
$\gamma$ has a different value. 

\begin{figure}[htb]
\epsfysize=15\baselineskip
\centerline{\hbox{ \epsffile{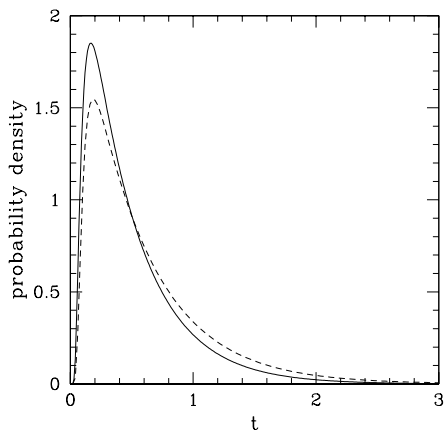}}}
\caption{\protect Probability distribution of
escape times measured in reduced, dimensionless, units.
The dashed curve corresponds to the logarithmic potential depicted 
in Fig.~\protect{\ref{potential}}, 
while the solid line is in the absence of the entropic barrier. 
The curves were obtained by numerical
integration of the Fokker--Planck equation.}
\label{esccomp}
\end{figure}

The problem of escape from a {\em deep} well in one dimension was 
considered by Kramers\cite{kramersphysica}. Assuming
that the escape rate is slow, i.e. if at every moment in time
the probability distribution of a particle in the well can be 
represented
by an equilibrium (Boltzmann) weight, Kramers' method enables 
an analytic calculation of the mean escape time. 
Applying Kramers' formula to the logarithmic potential of the problem 
one 
finds\cite{sung,muthu} (for $\gamma={1/2}$) that the mean escape time 
$\tau$ is $(\pi^2/16) N^2/D$. 
The distribution of  escape times in Kramers' formula is by 
construction
a simple exponential.  
Equation~\ref{stdeqn} can also be solved numerically, by placing
a delta function at the left edge of the potential depicted in 
Fig.~\ref{potential} at $t=0$ and integrating in time. 
There are some differences between  the numerical solution to 
Eq.~\ref{stdeqn},
and the Kramers' solution, e.g.
the former decays to zero for $t\rightarrow 0$ due to the time 
it takes for the delta function to diffuse out of the well.  
However, the mean escape time for the numerical solution
is $\tau\approx0.6N^2/D$ which almost coincides 
with the approximate Kramers' result. 
The distribution of escape times calculated from integrating the 
Fokker-Planck
equation is shown by the dashed line in Fig.~\ref{esccomp}.

If the  potential barrier is absent  altogether, then the distribution 
of 
escape times can again be calculated (e.g. by numerical integration), 
and is
depicted by the solid line in Fig.~\ref{esccomp}. 
The mean escape time in this
case is $\tau\approx 0.5$ (in reduced units of $N^2/D$). 
There is only a 20\% difference between the mean escape times of the 
problems with and without the  logarithmic potential barrier. 
More strikingly, there
is little difference in the distribution of times with or without a 
barrier.
(This barrierless version of the problem can be thought of as 
describing 
the adsorption of a particle starting at a unit distance from a sink. 
 From a polymer perspective, it corresponds to passage 
of a polymer through  a {\em ring} -- although the polymer
must pass through a constricted space, its free ends can 
have any possible configuration.)
In view of the minute effect of the entropic potential, the notion of 
`escape over a barrier' does not provide a particularly useful analogy.

A central assumption in the reduction of the polymer problem
to a single coordinate is that  translocation is slow.
Specifically, it should be slow enough that the polymer segments 
on the two sides of the membrane are in equilibrium at every 
value of $s$. 
We can  check for the self-consistency of this assumption:
The equilibration time of a free polymer can be 
estimated\cite{deGennes_book} as the time
required for it to diffuse a distance equal to its
own radius of gyration
$R_g$. Under Rouse dynamics (which ignores hydrodynamic effects), the
diffusivity of the center of mass of an $N$--monomer polymer is reduced 
to $D/N$, where $D$ is the diffusion constant of a single monomer,
resulting in an equilibration time of order of $R_g^2N/D$. 
For phantom polymers $R_g^2\sim N$, and, consequently the equilibration
time is of the same order ($N^2/D$) as the mean passage time
obtained with Brownian translocation dynamics. 
Thus the equilibrium assumption is marginally  (in)valid in this case. 
On the other hand, for self--avoiding polymers $R_g\sim N^\nu$, where 
$\nu=0.75$ and 0.59 for dimensions $d=2$ and $3$, respectively. 
The resulting relaxation times ($\sim N^{1+2\nu}/D$) are now
longer than the translocation times predicted ($\sim N^2/D$) by  
Brownian dynamics, and consequently the approximations involved
are not self-consistent. 
In the following sections we try to gain further insights into the 
problem
by numerical simulations of the translocation of a polymer in one and 
two
dimensions.

\section{Simulations in One--Dimension}\label{sim1d}

We begin by examining the translocation of a {\em one dimensional 
phantom\/}
polymer via a Monte Carlo simulation with Rouse-like dynamics as 
follows.
Our model consists of a chain of $N$ `atoms' placed
on the sites of a one--dimensional lattice. No 
excluded volume interactions are present, and the spatial distance 
between two neighboring atoms (along the sequence of the chain)
can be 0, 1 or 2 lattice spacings, i.e. the ``bond'' between
adjacent atoms has a maximal length of 2. This represents a trivial
implementation of the fluctuating--bond method\cite{carm}.
Initially, the first atom of the chain is
placed on, say, the right of the membrane, while all other atoms are on
the left. (We assume that the membrane is located between the
coordinate $x=0$ and $x=1$ and, consequently, the first atom is 
initially
placed at $x=1$ while the rest of the atoms are at $x\le0$.)
During the simulation the first atom of the chain is never 
allowed to move to the left of the membrane. 
The ``width'' $w$ of the hole is adjusted by changing the maximal 
number 
of bonds allowed to be simultaneously present at the hole.
An elementary move consists of randomly picking an atom and
attempting to move it one lattice step in a randomly selected 
direction.
If the new configuration does not violate any of the restrictions of
the model it is accepted. $N$ elementary atom move attempts are defined
as one Monte Carlo time unit.

\begin{figure}[htb]
\epsfysize=15\baselineskip
\centerline{\hbox{ \epsffile{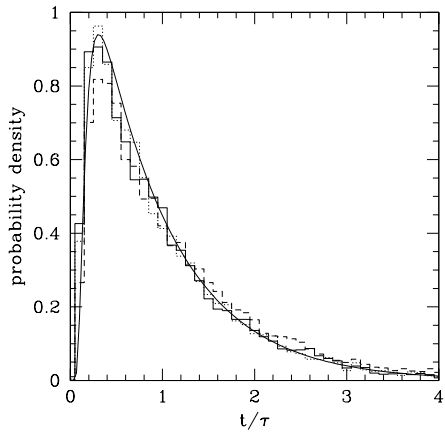}}}
\caption {\protect Probability density of translocation times
for one--dimensional phantom polymers. Solid, dotted and dashed lines
correspond to polymer lengths $N=32$, 64 and 128, respectively, and 
were extracted from 10,000 simulations, each. 
The times have been normalized by their respective mean translocation 
time $\tau(N)$ in each case. The continuous curve corresponds to 
the solution of the Fokker--Planck
equation in the single--particle approximation.}
\label{timdist}
\end{figure}

\begin{figure}[htb]
\epsfysize=15\baselineskip
\centerline{\hbox{ \epsffile{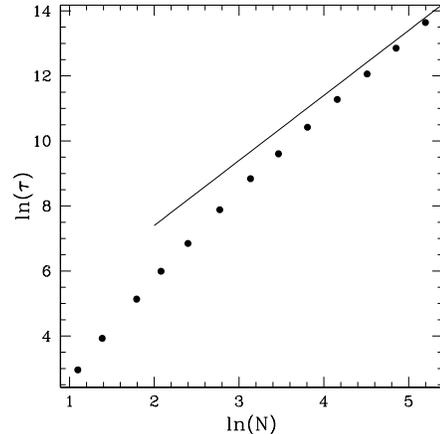}}}
\caption {\protect Logarithmic plot of the mean translocation time 
$\tau$ as
a function of chain length $N$ for a one dimensional phantom polymer.
Each data point represents an average over 10,000 realizations.
The solid line has slope 2.}
\label{TvsN}
\end{figure}

Each simulation is terminated when all monomers are on one side.
Figure~\ref{timdist} depicts the distribution of translocation
times of such phantom chains measured for several chain lengths,
and for unit width of the hole.
If we normalize the time for each $N$ by the mean translocation 
time $\tau$ for that length, we observe that the resulting curves are 
quite 
similar and closely resemble the theoretical curve obtained assuming a
Brownian translocation coordinate. 
While the similarity may appear to support this picture, 
we should note that since the distribution is
constrained to vanish at both short and long times, qualitative
similarities are dubious.
Moreover, the absolute values of the mean translocation time $\tau$,
as depicted in  Fig.~\ref{TvsN}, are significantly larger than the 
estimates from Brownian dynamics. 
This log--log plot indicates that the apparent exponent is somewhat 
larger 
than 2 for small $N$, and only gradually approaches the scaling 
form $\tau\sim N^2$. 
However, the prefactor of the power law for $N=256$ is roughly two 
orders of 
magnitude larger than expected for a Brownian translocation coordinate.
Such discrepancy should not be surprising -- the translocation
time predicted by this model are similar to the time required for 
a polymer to diffuse its own radius of gyration. 
It is reasonable to expect that passage through a
narrow hole should be slower than diffusion without a wall. 

\begin{figure}[htb]
\epsfysize=15\baselineskip
\centerline{\hbox{\epsffile{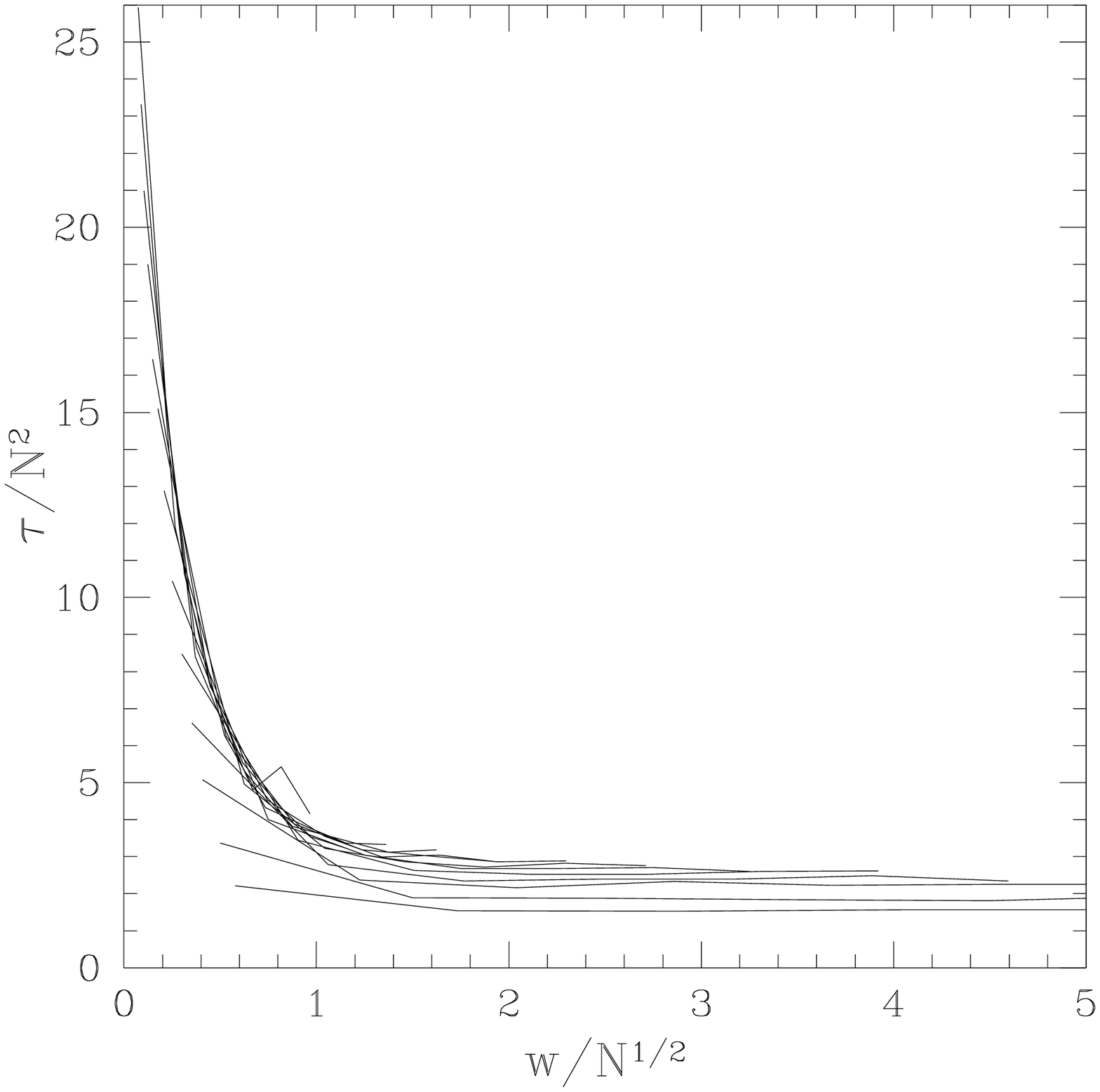}}}
\caption {\protect Dependence of the mean translocation time $\tau$ on 
the width of the hole $w$, for $N=3$, 4, 6, 8, $\dots$ 91, 128, and 
181.
The times have been normalized by $N^2$ to focus on the behavior
of the prefactor. Each curve corresponds to fixed $N$, and is obtained
by averaging over 1,000 cases.}
\label{TvsW}
\end{figure}

Not surprisingly, translocation times are strongly affected
by the width $w$ of the hole, as indicated in Fig.~\ref{TvsW}.
When $w$ reaches the size $R_g$ of the chain
($\sim \sqrt{N}$) the translocation time should become independent of 
$w$.
This is supported by the saturation of the rescaled times $\tau/N^2$ 
at roughly the same value of $w/R_g\sim w/\sqrt{N}$ in this figure.
For small holes, the translocation times are strongly dependent on the 
hole size, and  show some indications of collapse onto a universal 
curve, 
but the $N^2$ dependence is not as clear as for the wider holes.

\section{Simulations in Two Dimensions}\label{sim2d}

Excluded volume effects drastically modify the shape and
properties of the one--dimensional phantom polymer considered in the
previous section.
The chain becomes stretched, and its dynamics are then limited by 
reptation\cite{deGennes_book}. 
To study the effects of self-avoidance on translocation in the coiled
state, higher dimensional simulations are necessary.
Two--dimensional polymers are ideally suited to this purpose
for the dual reasons that excluded volume effects are more apparent,
while computation times are shorter than in the three--dimensional 
case.

As in the 1-d system, we employ a fluctuating--bond model for 
simulations\cite{carm}, implementing Rouse-like dynamics for linear 
polymers
of several different lengths $N$. Simulations were performed both with 
and 
without excluded volume constraints. In this model, the monomers of the 
polymer lie on a 2-d square lattice. Random motion is simulated through 
a 
series of elementary moves of single monomers. In each move, a monomer 
is 
selected randomly and then moved a single lattice unit in one of the 
$+x, -x, +y,$ or $-y$ directions. If the move violates any of several 
constraints, it is rejected. For phantom chains, bonds have a maximal 
allowed length of $\sqrt{10}$ lattice units. For excluded volume 
chains, 
in addition, the distance between any two monomers is constrained to be 
at 
least $2$ lattice units. For excluded volume chains, these constraints 
also prevent the chain from crossing itself\cite{carm}. 
The wall has a thickness of $3$ units and the 
hole has a width of $2$ lattice units. At these sizes, only one monomer 
may be
in the hole at a time, but the hole is large enough that translocation 
can 
occur with the given move set. Each $N$ elementary atom move attempts 
are 
defined as one Monte Carlo time unit.

The simulation begins by placing the first monomer at the hole, 
while the remaining $N-1$ monomers are in a random conformation on the 
left side of the wall. (To generate the initial random configuration, 
the chain is first allowed to fluctuate subject to the constraints of 
impermeability of the wall and fixed location of the first monomer.
The fluctuation time is $20N^2$, which should be sufficient to 
randomize 
the initial condition for the purposes of our 
simulations, at the length scales employed. In any case, the time 
it takes for translocation is many times longer than the Rouse 
relaxation 
time, and any initial condition effects will be transients.) 
Once the initial configuration has been
established, the polymer is allowed to move in accordance with
the restrictions of the model. We measure the time between the
beginning of the translocation, and the moment when the last monomer
enters the hole. Because only one monomer can lie in the aperture 
at a time, this condition is equivalent to the complete translocation 
of the 
polymer through the hole. As in the one-dimensional case, the first 
monomer 
is not allowed to move to the left of the hole. 

\begin{figure}[htb]
\epsfysize=15\baselineskip
\centerline{\hbox{\epsffile{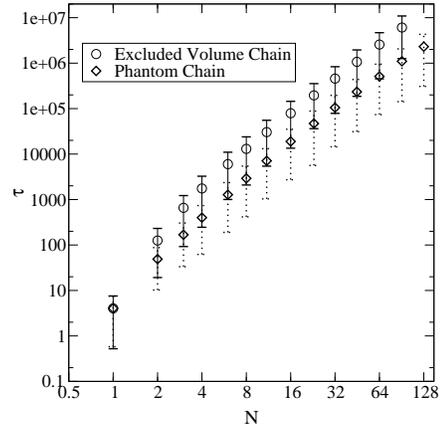}}}
\caption {\protect Logarithmic plot of the mean translocation time
$\tau$ as a function of chain length $N$, for two-dimensional phantom 
and
excluded volume polymers. Error bars indicate the standard deviation 
over
runs.}
\label{2dTransTimes}
\end{figure}

Translocation times were calculated for a number of
different chain lengths $N$, with several thousand runs at each length. 
(The number of runs decreases with increasing $N$, due to 
CPU limitations.) Theses results are shown for both phantom and
self-avoiding chains in Fig.\ref{2dTransTimes}. Error bars indicate the
standard deviation of translocation times over runs.

\begin{figure}[htb]
\epsfysize=15\baselineskip
\centerline{\hbox{\epsffile{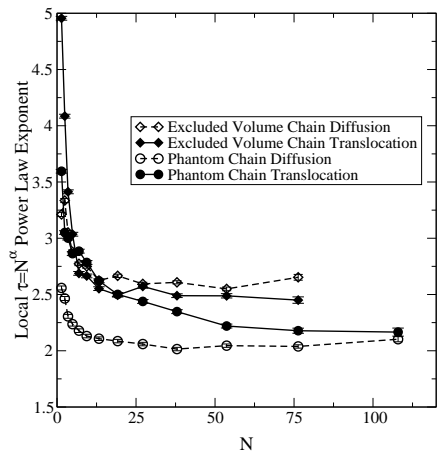}}}
\caption {\protect The effective power law exponents  for translocation 
and 
diffusion times as a function of length $N$ for 2-d polymers. 
For both translocation and free diffusion, self-avoiding chains 
exhibit a  dynamic exponent that approaches 2.5, equal to $1+2\nu$ 
given 
the $d=2$ swelling exponent of $\nu=3/4$. 
For phantom chains, the exponent is noticeably smaller, and approaches 
2. 
In both the self-avoiding and phantom cases, the asymptotic exponent 
for 
translocation time is nearly equal to that for Rouse relaxation.}
\label{2dPowerLaw}
\end{figure}

We then attempted to fit the results to a dynamic scaling form, with 
the
effective exponent depicted in Fig.~\ref{2dPowerLaw}. 
The points are a plot of the local exponent $\alpha$, assuming a power 
law 
$\tau \sim N^\alpha$, as determined from two neighboring polymer 
lengths 
$N_1$ and $N_2$ via the formula 
$\alpha(\sqrt{N_1 N_2}) = \log [\tau(N_2)/\tau(N_1)]/ \log(N_2/N_1)$.
The  exponent $\alpha$ for the excluded volume  translocation 
approaches 
an asymptotic limit which is definitely larger than 2. 
This clearly indicates that excluded volume effects are inconsistent 
with Brownian dynamics for the translocation coordinate. 
The data for the translocation times of a phantom polymer (in which the 
self-avoiding constraint of the bond-fluctuation model is removed)
are also included in this figure. 
The power law exponent in this case asymptotically  approaches a value 
close to 2, in agreement with the one dimensional results, and 
consistent
with Brownian dynamics. At each $N$, the uncertainty in the average 
translocation time (equal to the standard deviation of $\tau(N)$ over
runs divided by the square root of the number of runs) 
has been translated into an uncertainty in the local power law.  
This uncertainty, as depicted by the error bars, is quite small, 
indicating that our conclusion is not due to
statistical fluctuations.

The observed scaling laws for self-avoiding and phantom polymers in 
fact agree 
with the exponents expected theoretically for Rouse relaxation in the 
absence of a wall. 
As stated previously, translocation across the
barrier requires that, minimally, the chain diffuses a distance 
equal to its radius of gyration. The time for such diffusion is 
$\tau_R \sim R_g^2N/D \sim N^{1+2\nu}$. For a self-avoiding
walk in 2 dimensions, $\nu = 3/4$ and $\tau_R \sim N^{2.5}$, while for 
a phantom chain, $\nu$ is replaced by $1/2$, leading to the 
relationship 
$\tau_R \sim N^2$.
For comparison, Fig.~\ref{2dPowerLaw} also shows the derived power law
exponents from simulations of simple diffusion in the absence of a 
wall. 
For an excluded volume chain, we approach the scaling $\tau \sim 
N^{2.5}$,
while for the phantom chain, $\tau \sim N^2$.
The latter results were generated as follows: We started with the same
type of random initial configurations (generated by annealing near an 
impermeable wall) as for the translocation case, and again imposed
the constraint that  the first monomer cannot pass to the left side of 
an imaginary wall. The other monomers, however, are allowed to 
diffusive
without feeling this wall.
The simulation was stopped when all monomers moved to the right of the
imaginary wall.

\begin{figure}[htb]
\epsfysize=15\baselineskip
\centerline{\hbox{\epsffile{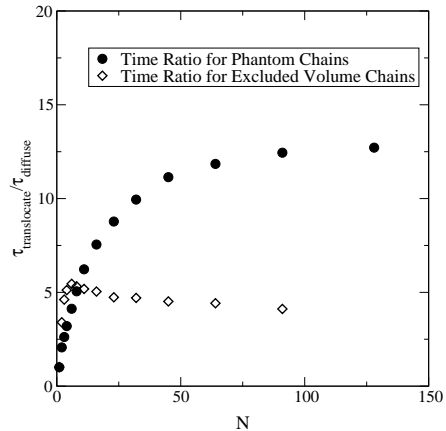}}}
\caption {\protect
Comparison of simulation results for diffusion through 
a hole that can fit one monomer 
versus free diffusion.
Translocation times through the hole  are larger 
than the free diffusion for
both phantom and self-avoiding chains. In both cases, the ratio of 
these times
levels off to a constant.
}
\label{2dHoleEffect}
\end{figure}

At least in these examples, we find that the scaling of translocation 
times
is the same as that for equilibration of the polymer in the absence of 
a 
wall. However, the constraint of passing through the hole must clearly 
slow
down the dynamics of the polymer compared to the case of free diffusion
over a similar distance.
This slowdown must then be reflected in an overall prefactor which 
determines
how much slower translocation is relative to pure diffusion.
The data that address this issue are displayed in 
Fig.~\ref{2dHoleEffect}. 
The ratio of crossing times for translocation compared to  pure 
diffusion 
(as described in the previous paragraph) are plotted for both  
self-avoiding and phantom chains. 
For self-avoiding polymers, the hole slows down the chain by a factor 
of about $5$; while for phantom chains the ratio is about $13$. 
These numbers 
are roughly $N$-independent for $N>10$ (for excluded volume chains) and
$N>50$ (for phantom chains).
It is interesting to note that translocation slows down the phantom 
chain more than it does the excluded volume chain. 
However, these ratios should not be taken too seriously, since for
more realistic systems such as the translocation of DNA molecules 
through 
a cell membrane, the details of the shape and interaction forces at the 
pore 
play a significant role \cite{Kasianowicz,Lubensky,meller}.

\section{Discussion}\label{discussion}

The central result of this paper is the nontrivial scaling of the mean 
translocation time $\tau$ for unforced passage of a polymer through a 
hole with its length $N$.
For the diffusive dynamics of a self-avoiding chain in $d=2$, numerical
simulations indicate $\tau\sim N^{2.5}$.
By extension, we expect $\tau\sim N^{\alpha}$ with $\alpha=1+2\nu$
for diffusive dynamics.
Possibly even more generally for other types of dynamics $\alpha=\nu 
z$,
where the swelling exponent $\nu$ relates the radius of 
gyration of the polymer to its length by $R_g\sim N^\nu$, 
while its relaxation time scales as $\tau_r\sim R_g^z$ with the dynamic 
exponent $z$.

The natural parameter for following the progress of the translocation 
process
is the label $s(t)$ for the monomer in the hole at time $t$ 
(see Fig.~\ref{transloc}).
A commonly used analogy is that this `reaction coordinate' undergoes
stochastic motion, as in a Brownian particle in a force field
\cite{muthu,park,Lubensky}.
For unforced translocation (with or without inclusion of an entropic
barrier), such approaches lead to $\tau\sim N^2$, which
is inconsistent with our numerical results, 
except for the case of diffusing phantom 
polymers. 
Such scaling is also at odds with the expectation that constraining the
polymer to pass through a hole should slow down its dynamics compared 
to
free motion over a similar distance. 
The numerically observed scaling is thus indicative of 
{\em anomalous dynamics} of the translocation coordinate $s(t)$ ---
we propose instead an alternative analogy that incorporates this 
observation.

The dynamics of the polymer is the result of cooperative motions of its
many monomers. 
When described by just a single variable, say the translocation 
coordinate,
the effect of the other degrees of freedom is to exert correlated 
forces
leading to anomalous dynamics.
This is best illustrated by the well studied case of the dynamics of a
single monomer in a polymer:
Consider the position ${\vec r}(t)$ of a particular atom in the
background of all the other monomers.
The  dynamics of ${\vec r}(t)$ has indeed been studied by 
numerical, analytical, and experimental methods\cite{KBGC};
its anomalous features are easily captured by the following scaling 
argument:
For short times, we expect that the squared change in the position
has a scaling form
$\left\langle \Delta r^2(t)\right\rangle\sim t^{2\tilde\zeta}$,
with no dependence
on $N$ since the monomer does not yet feel any effects from the
finite size of the polymer.
At a time of the order of the relaxation time $\tau_r$ for the whole 
chain,
the monomer should have moved by a distance of the order of its radius
of gyration, giving $R_g^2\sim \tau_r^{2\tilde\zeta}$. 
Since $\tau_r\sim R_g^z$, we immediately obtain the exponent
$\tilde\zeta=1/z$ describing the anomalous fluctuations of the 
specified
monomer at short times.

We now adapt a similar scaling argument to describe the 
squared change in the translocation 
coordinate, assuming $\left\langle \Delta s^2(t)\right\rangle\sim 
t^{2\zeta}$
at short times. This behavior should saturate when $s$ becomes of the 
order
of the chain length in a time $\tau\sim R_g^z\sim N^{\nu z}$
(assuming that translocation times always scale in the same way as 
equilibration times).
Substituting this in the former equation allows us to identify the 
exponent
$\zeta=1/(z\nu)$ for anomalous dynamics of the translocation 
coordinate.
For the case of diffusive polymer dynamics, we thus obtain 
$\zeta=1/(1+2\nu)$
resulting in $\zeta=2/5$ and $\zeta\approx 0.46$ for self-avoiding
chains in two and three dimensions respectively;
i.e. in this case the fluctuations are {\em subdiffusive}.
If we naively use the Zimm exponent $z=d$ describing relaxation of 
polymers
in hydrodynamic flows in $d$ dimensions, we obtain
$\zeta(d=2)=2/3$, while $\zeta(d=3)\approx 0.56$, i.e. the fluctuations
are predicted to be {\em superdiffusive} in this case.
The usual origin of the speed up of polymer dynamics in a fluid is
attributable to the velocity flow field set up by the other monomers.
It is doubtful that using the bulk Zimm exponent remains valid for 
flows 
which 
must vanish in the vicinity of the wall. Thus the above 
prediction of
superdiffusive behavior should not be taken seriously prior to a
proper analysis 
of the hydrodynamic correlations in the vicinity of 
the wall.

It is experimentally hard to directly probe the motion of the 
translocation
coordinate.
The quantity that is easily measured in experiments\cite{meller} is
the distribution of the translocation times {\em in the presence of a 
force}
introduced via a chemical potential difference $\Delta\mu$ for monomers 
on the
two sides of the wall.
While our results so far were in the absence of such a driving force,
anomalous dynamics has consequences for the length dependence of the 
forced velocity.
Let us first recall the arguments for the drift velocity $v$ of a 
polymer in 
a force $\vec F$: Scaling considerations suggest 
$v(F)\sim (R_g/\tau_r)\phi(FR_g/k_BT)\sim N^{\nu(1-z)}\phi(FN^\nu)$,
where $\phi$ is a scaling function depending on the ratio of two
quantities having dimensions of energy.
The proportionality of the velocity to the force requires a linear 
scaling
function, leading to a mobility $v/F\sim N^{-\nu(z-2)}$.
For Rouse and Zimm dynamics this leads to the well-known scalings of 
mobility
as $1/N$ and $1/R_g^{d-2}$, respectively.
Similarly, for the pulling velocity $u\equiv \dot s$ of the 
translocation 
coordinate,
scaling suggests $u\sim N/\tau\phi(\Delta\mu N/k_BT)\sim 
N^{2-z\nu}\Delta\mu$.
Only for the case of diffusive dynamics of a phantom chain this 
velocity
is independent of $N$. 
The anomalous slowdown due to Rouse dynamics leads to a mobility that 
scales
as $N^{-1/2}$ in $d=2$, and $N^{-0.18}$ in $d=3$.
By contrast, hydrodynamic speeding up with Zimm dynamic exponents leads 
to
a mobility that grows as $N^{1/2}$ and $N^{0.23}$ in two and three 
dimensions
respectively. Once more, the latter results are not to be taken 
seriously
without full hydrodynamic calculations in the presence of a wall.
Experiments so far\cite{meller} do not indicate anomalous scaling, but
the size range may not be sufficient to detect the rather small 
exponent.

We conclude by listing two other avenues of potential exploration.
The first is to note that the distribution of translocation times 
should also
be modified by the anomalous dynamics, potentially to include power law 
tails 
that are distinct from the exponential tails in
Fig.~\ref{esccomp}~\cite{metzler,ding}. Secondly, it may be possible to
construct an experimental system close to our two dimensional 
simulations,
using vibrated granular chains in a variation on the set up used in
Ref.~\cite{ben-naim}, with a chain that is threaded through a hole in a 
wall.

\acknowledgements 
We thank M. Muthukumar, K. Kremer, E. Ben-Naim, and Z.A. Daya for 
useful 
discussions. This work was supported by the US--Israel Binational
Science Foundation Grant No. 1999-007, and  by the National Science 
Foundation 
through grants No. DMR-01-18213, and PHY99-07949 (M.K.).

\end{multicols}

\begin{thebibliography}{99}
\bibitem{sung-comment} In Ref.\cite{sung} it is noted that the 
parameter $D$ 
can
depend on the monomer number, and that if taken as a constant, it 
should scale
with the number of monomers $N$.  However, we could not reproduce the
proposed dependence on $N$.
\end{thebibliography}
\end{document}